\documentstyle[aps,12pt,epsf]{revtex}
\begin{document}
\title{A soft mode in a finite Fermi-system:\\
anharmonic effects near the instability point}
\author{Vladimir Zelevinsky$^{1,2}$ and Alexander Volya$^2$}
\address{
\baselineskip 14pt
$^{1}$Department of Physics and Astronomy and\\
$^{2}$National Superconducting Cyclotron Laboratory,\\
Michigan State University, East Lansing, Michigan 48824-1321 USA}
\maketitle 
\vskip 10 pt
\begin{abstract}
We consider a finite Fermi-system where the residual interactions create a soft 
mode of the excitation spectrum. Because of the large vibrational amplitude,
the standard random phase approximation does not work in this situation. We
develop a regular method for constructing the anharmonic potential
and illustrate the application of the formalism by a simple model.
\baselineskip 14pt
\end{abstract}

\baselineskip 14pt
\section{Introduction}
The appearance of a soft collective mode of vibrational nature is quite common 
in nuclei and in other mesoscopic quantum systems, such as atomic clusters.
The conventional theoretical way to handle this situation 
\cite{bel,BE,ring,solov}
goes along the standard line: mean field (MF) $-$ residual interactions $-$ 
random phase approximation (RPA). The MF determines the symmetry of the system
around the ground state and corresponding elementary excitations, fermionic
quasiparticles. The residual interactions 
include both coherent effects and collision-like processes
responsible for the chaotization of motion and lead from a Fermi-gas to 
Fermi-liquid. The formation of coherent modes is
described within the framework of the RPA where the quanta of those modes
are treated as independent quasibosonic excitations \cite{BE,solov}. 
In the low-lying states, 
kinematic corrections due to the fact that the quanta are built of fermions,
as well as high order dynamic effects, can be accounted for perturbatively 
\cite{BE,KM}. 

The RPA-type theories become insufficient when, in some region of the 
parameters, the vibrational frequency $\omega$ approaches zero.
The vibrational amplitude then grows $\propto 1/\sqrt{\omega}$
revealing the instability of the MF. In a finite system, 
this is not necessarily a vestige of a phase transition or sharp restructuring 
of the system. Rather it might be a signature of the failure of the 
theoretical consideration based on the picture of harmonic vibrations. For
instance, a low energy of the first excited quadrupole state does not mean
that the nucleus becomes deformed. In the regime of large amplitude
collective motion, we need to reject the harmonic approximation and find a
way of calculating the anharmonic effects which cannot be here treated as
small corrections. 

The estimates \cite{lim}, as well as more detailed
unrestricted MF calculations, show that in the quadrupole case the effective 
potential is close to the $\gamma$-unstable \cite{gamm} quartic one, 
$\sim \beta^{4}$, and the spherical symmetry of the MF still holds but only 
on average. The popular interacting boson model \cite{IBM} with the phonon
number fixed by a number of fermion pairs cannot describe the soft vibrational
bands which stretch to very high spins. In this sense the
phenomenological models \cite{quar,VZ,arm,iach} based on a specific form of 
the collective quadrupole Hamiltonian are more successful. The onset of 
deformation occurs beyond the point of the RPA instability when some previously
unoccupied configurations sharply lower their energy as a function of 
deformation and thereby select the equilibrium static value of $\gamma$, 
usually $\gamma=0$. The macroscopic analog of this scenario would be the first 
order phase transition.

Below we consider a typical collective soft mode and show a way of 
constructing the effective nonperturbative anharmonic potential for 
large-amplitude collective motion starting with the full microscopic many-body
Hamiltonian. We present only the skeleton of the formalism and apply it to the
Lipkin model known as a testing ground of various theoretical approximations
(the first version of this approach was published in \cite{sol}).

\section{Generalized density matrix}

It is convenient to use the operator language with the generalized 
density matrix (GDM) as the main tool \cite{denmat,nucold,supp,MF}. We consider 
a truncated single-particle space of orbitals $|1)$ of full dimension $\Omega$.
The GDM is the set of the operators 
\begin{equation}
R_{1\, 2}=a^{\dagger}_2\,a_{1}\,,                         \label{form_1}
\end{equation} 
where $a$ and $a^{\dagger}$ are fermionic operators in the second quantization
(in a similar way, one can consider Bose-systems), the subscripts 
{\sl 1,2} form a matrix in single-particle space $\Omega$, and each element 
$R_{1\,2}$ is an operator in the many-body Hilbert space. These operators 
generate a closed $SU(\Omega)$ Lie algebra given by the commutation relations
\begin{equation}
[R_{1\, 2},\,R_{3\, 4}]=\delta_{1\,4}\,R_{3\,2}-\delta_{2\,3}\,R_{1\,4}\,;
                                                        \label{form_algebra}
\end{equation}
the trace of the GDM in single-particle indices (tr) gives a number operator, 
${\rm tr}(R)=N$. 
The GDM is Hermitian in the combined space of single-particle and
many-body variables, $R_{1\,2}^{\dagger}= R_{2\,1}$.

The dynamics of the system are governed by a standard Hamiltonian which
contains one-body and two-body terms,
\begin{equation}
H=\sum_{1,\,2}\,\epsilon_{1\,2} a_1^{\dagger} a_2 + \frac{1}{4}\,
\sum_{1,\,2,\,3,\,4}\, V_{1\,2;\,3 4}\, a^{\dagger}_1 a^{\dagger}_2
a_3 a_4\,,                                                   \label{form_2}
\end{equation}
where $\epsilon_{1\,2}=\epsilon_{2\,1}^{\ast}\,,\quad V_{1\,2;\,3\,4}=
V_{4\,3;\,2\,1}^{\ast}$,
and we assume the antisymmetrized form of the two-body interaction.
We define a self-consistent field $W$ (similarly to $R$, an operator in the
combined space) as a linear functional of the GDM,
\begin{equation}
W_{1 4}\{R\}\equiv \frac{1}{2}\,\sum_{2,\,3}\,V_{1\,2;\,3 4}\,R_{3\,2}=
\frac{1}{2}\,\sum_{2,\,3}\,V_{1\,2;\,3 4}\,a^{\dagger}_{2}\,a_{3}\,. \label{e2}
\end{equation}
The hamiltonian in eq. (\ref{form_2}) can also be written in terms of the GDM.

The equations of motion for the creation and annihilation
fermionic operators are 
\begin{equation}
[a_1,\,H]=\sum_{2}\,(\epsilon_{1\,2}+W_{1\,2})\,a_{2}\,,\quad
[a^{\dagger}_1,\,H]=-\sum_{2}\,a_{2}^{\dagger}\,(\epsilon_{2\,1}
+W_{2\,1})\,,                                                     \label{e3}
\end{equation}
whereas the total GDM (\ref{form_1}) satisfies the nonlinear operator equation
\begin{equation}
[R,\,H]=[\epsilon+W\{R\},\,R]\,.                     \label{form_motion}
\end{equation}
These equations are still exact.
Here the commutators are understood to act in the combined space, for example,
\begin{equation}
[W,R]_{12}\equiv \sum_{3}(W_{13}R_{32}-R_{13}W_{32});     \label{e4}
\end{equation}
all elements are many-body  operators. 

\section{Mapping onto collective space}

Now we make two crucial assumptions: (i) there exists a ``collective band" as a
set of stationary states which are coupled by strong intraband transition 
amplitudes while the transitions to the states of a different nature are weak
and can be ignored, or be taken into account perturbatively later; (ii)
the nomenclature (quantum numbers) of the band states can be built with the 
aid of the operators of collective coordinates $\alpha$ and conjugate momenta 
$\pi$. These assumptions are fulfilled accurately \cite{jolos}
for low-lying
quadrupole vibrations in medium and heavy spherical nuclei where it is known
that the quadrupole transitions from the ground state are nearly
saturated by the first excited $2^{+}$ state, which in turn gives rise to
transitions to the ``two-phonon" triplet of states $0^{+}, 2^{+}$ and $4^{+}$,
and so on. This means that there is a good correspondence between the ideal
quadrupole phonon space and realistic spectra in spite of the fact that the
predictions of the naive model of harmonic quadrupole vibrations are badly
violated. If so, the observed states can be generated by the quadrupole
coordinate and momentum operators $\alpha_{2\mu}$ and $\pi_{2\mu}$ although the
collective Hamiltonian $H(\alpha,\pi)$ might be very far from the harmonic one.

According to our assumptions, the collective subspace is spanned by the
operators $\alpha$ and $\pi$ with normal commutation relations
\begin{equation}
[\alpha,\,\pi]=i\,,                                           \label{e5}
\end{equation} 
and their high order products. For simplicity we take here scalar quantities;
the rotational tensor character can be introduced in a straightforward way.
The general form of the effective Hermitian 
collective Hamiltonian acting within this subspace is
\begin{equation}
{\cal H}=\sum_{m,\,n}\,\frac{\Lambda^{(m\, n)}}{2 m n}\,
[\alpha^m,\,\pi^{n}]_{+}.                                   \label{sm_ham}
\end{equation} 
Our goal is to derive the unknown $c$-number coefficients $\Lambda^{(mn)}$
from the microscopic Hamiltonian $H$, eq. (\ref{form_2}).
This can be done by the corresponding mapping of the exact 
operator equations of motion (\ref{form_motion}).

We are interested in the matrix elements of the equations of motion between
collective states. Since the dynamics are saturated in the collective space, we
leave as the intermediate states in those equations only the states 
within the band. Then operators $R$ and $W$ can be effectively represented
by the functions of $\alpha$ and $\pi$ similarly to (\ref{sm_ham}),
\begin{equation}
{\cal R}=\sum_{m,\,n}\,\frac{r^{(m\, n)}}{2 m n}\,[\alpha^m,\,\pi^{n}]_{+}, 
\quad
{\cal W}=\sum_{m,\,n}\,\frac{w^{(m\, n)}}{2 m n}\,[\alpha^m,\,\pi^{n}]_{+}\,.
                                                            \label{sm_GDM}
\end{equation} 
The question of mapping is now formulated as a problem of finding a set of 
numbers $\Lambda^{(m\, n)}$ and quantities $r^{(m\, n)},\,w^{(m\,n)}$  
(matrices in single-particle space) which express the contributions of 
specific elementary excitations to a given collective operator. The
corresponding parts of $r^{(m\,n)}$ and $w^{(m\,n)}$ are interrelated by the
self-consistency conditions (\ref{e2}).

Physical arguments of time-reversal (T) invariance and the possibility of 
canonical transformations in the
collective space, such as shifts and rescalings of
collective variables, allow us to consider only $\Lambda^{(m\, n)}$
with even $n$ and start the sum with the harmonic terms, $(mn)=(02)$ and (20),
so that eq. (\ref{sm_ham}) becomes ($\Lambda^{(0\,2)}=1/B, \,\Lambda^{(2\,0)}
=C$ give the mass and force parameters of the harmonic part)
\begin{equation}
{\cal H}=\frac{1}{2B}\pi^2+ 
\frac{C}{2}\alpha^2+{\Lambda^{(3\,0)}\over 3}\alpha^3+
{\Lambda^{(1\,2)}\over 4}[\alpha,\pi^2]_{+}+{\Lambda^{(4\,0)}\over 4}\alpha^4+
{\Lambda^{(0\,4)}\over 4}\pi^4 + {\Lambda^{(2\,2)}\over 8}[\alpha^2,\pi^2]_{+}
+\dots\,.                                                \label{sm_ham1}
\end{equation}    

Under our assumptions, the full operator equations of motion
(\ref{form_motion}) should be satisfied inside the band. Therefore we require
that in this space
\begin{equation}
[{\cal R},\,{\cal H}+\epsilon+{\cal W}\{{\cal R}\}]=0\,.      \label{sm_eq1}
\end{equation}
Commutators involving ${\cal H}$ and $\epsilon$ in the above expression are 
simple since ${\cal H}$ does not contain single-particle variables,
whereas $\epsilon$ is a $c$-number matrix in the Hilbert space. The 
commutator of ${\cal W}$ with the GDM is more complex as both operators act in 
the combined space. 

Below we show the lowest order equations. As seen from (\ref{sm_eq1}), 
it is convenient to introduce a self-consistent MF Hamiltonian as 
\begin{equation}
h=\epsilon+{\cal W}\{\rho\}\,,                                     \label{e6}
\end{equation} 
and a self-consistent RPA operator $\hat{L}$ defined \cite{nucold,supp}
by its action on an arbitrary single-particle matrix $r$,
\begin{equation}
\hat{L}r=[h,r]+[{\cal W}\{r\},\,\rho]\,.                         \label{e7}
\end{equation}
In eqs. (\ref{e6}) and (\ref{e7}) we used the ground state single-particle 
density matrix $\rho\equiv r^{(0\,0)}$. The lowest static part, $(mn)=(00)$,
produces a set of the MF equations
\begin{equation}
0=\left [ h,\,\rho \right ]+ i \delta^{(0\,0)}\,,       \label{form_static}
\end{equation}
where $\delta^{(0\,0)}$ is a correction to the usual Hartree-Fock approximation
from higher orders which changes the 
average MF single-particle occupancies (eigenvalues of $\rho$) due to the 
fluctuation effects coming from the soft mode \cite{MF}.
The next set of equations corresponds to the parts linear in $\alpha$ and $\pi$ 
operators (T-even and T-odd terms, respectively),
\begin{equation}
-i \Lambda^{(2\,0)} r^{(0\,1)}=\hat{L} r^{(1\,0)}+ i \delta^{(1\,0)}\,,
                                                         \label{form_rpa1}
\end{equation} 
\begin{equation}
i \Lambda^{(0\,2)} r^{(1\,0)}=\hat{L} r^{(0\,1)}+i \delta^{(0\,1)}\,.
                                                       \label{form_rpa2}
\end{equation}
These terms are analogous to the RPA although it is not assumed that the
occupation numbers are 0 and 1.
The following three equations in quadratic order are 
\begin{equation}
-i \Lambda^{(2\,0)} r^{(1\,1)}-i \Lambda^{(3\,0)} r^{(0\,1)}=
\frac{1}{2}\,\hat{L} r^{(2\,0)}+
[w^{(1\,0)},r^{(1\,0)}]+i \delta^{(2\,0)}\,,        \label{form_20}
\end{equation}
\begin{equation}
-i \Lambda^{(2\,0)} r^{(0\,2)}+i \Lambda^{(0\,2)} r^{(2\,0)}+ 
i \Lambda^{(1\,2)} r^{(1\,0)}=
\hat{L} r^{(1\,1)}+[w^{(1\,0)},r^{(0\,1)}]+[w^{(0\,1)},r^{(1\,0)}]
+i \delta^{(1\,1)}\,,                                \label{form_11}
\end{equation}
\begin{equation}
i \Lambda^{(0\,2)} r^{(1\,1)}-i \Lambda^{(1\,2)} r^{(0\,1)} =
\frac{1}{2}\,\hat{L} r^{(0\,2)}+
[w^{(0\,1)},r^{(0\,1)}]+i \delta^{(0\,2)}\,.       \label{form_02}
\end{equation}
We limit ourselves here to the fourth order of anharmonicities, i.e. cubic 
operators in equations of motion. The four corresponding equations are
\begin{eqnarray}
\nonumber
-\frac{i}{2} \Lambda^{(2\,0)} r^{(2\,1)}-i \Lambda^{(3\,0)} r^{(1\,1)}- 
i \Lambda^{(4\,0)} r^{(0\,1)}=
\frac{1}{3}\,\hat{L} r^{(3\,0)}+\frac{1}{2}\,[w^{(2\,0)},r^{(1\,0)}]+
\frac{1}{2}\,[w^{(1\,0)},r^{(2\,0)}]
+i \delta^{(3\,0)}\,,
\\
\nonumber
-i \Lambda^{(2\,0)} r^{(1\,2)}+i \Lambda^{(0\,2)} r^{(3\,0)}- 
i \Lambda^{(3\,0)} r^{(0\,2)}+i \Lambda^{(1\,2)} r^{(2\,0)}+
\frac{i}{2} \Lambda^{(2\,2)} r^{(1\,0)}=
\\
\nonumber
\frac{1}{2}\,\hat{L} r^{(2\,1)}+\frac{1}{2}\,[w^{(2\,0)},r^{(0\,1)}]+
[w^{(1\,1)},r^{(1\,0)}]
+\frac{1}{2}\,[w^{(0\,1)},r^{(2\,0)}]+[w^{(1\,0)},r^{(1\,1)}]
+i \delta^{(2\,1)}\,,
\\
\nonumber
-i \Lambda^{(2\,0)} r^{(0\,3)}+i \Lambda^{(0\,2)} r^{(2\,1)}+\frac{i}{2} 
\Lambda^{(1\,2)} r^{(1\,1)}-\frac{i}{2}\, \Lambda^{(2\,2)} r^{(0\,1)}=
\\
\nonumber
\frac{1}{2}\,\hat{L} r^{(1\,2)}+\frac{1}{2}\,[w^{(0\,2)},r^{(1\,0)}]+
[w^{(1\,1)},r^{(0\,1)}]
+[w^{(0\,1)},r^{(1\,1)}]+\frac{1}{2}\,[w^{(1\,0)},r^{(0\,2)}]
+i \delta^{(1\,2)}\,,
\\
\frac{i}{2} \Lambda^{(0\,2)} r^{(1\,2)}- \frac{i}{2} 
\Lambda^{(1\,2)} r^{(0\,2)}+ i \Lambda^{(0\,4)} r^{(1\,0)}=
\frac{1}{3}\,\hat{L} r^{(0\,3)}+\frac{1}{2}\,[w^{(0\,2)},r^{(0\,1)}]+
\frac{1}{2}\,[w^{(0\,1)},r^{(0\,2)}]
+i \delta^{(0\,3)}\,.
                                                 \label{sm_equation}
\end{eqnarray}
The higher order corrections $\delta^{(i,j)}$ arise from the commutators 
$[{\cal R},\,{\cal W}]$ and $[{\cal R}\,,{\cal H}]\,,$
\[\delta^{(0\,0)}={1\over 2}\left ( \left [w^{(1\,0)},\,r^{(0\,1)}
\right ]_{+}-\left [w^{(0\,1)},\,r^{(1\,0)}
\right ]_{+}+\dots \right),\]                     
\begin{equation}
- \left (\frac{1}{2}
\Lambda^{(2\,0)} r^{(1\,3)}-\frac{1}{2}\Lambda^{(0\,2)} r^{(3\,1)}+
\frac{2}{3}\Lambda^{(3\,0)} r^{(0\,3)}-\frac{1}{4}\Lambda^{(1\,2)} r^{(2\,1)}
+\dots \right )\,.                                \label{form_delta00}
\end{equation}
Each next term denoted by dots is four orders higher than the previous one. 
Furthermore, the lowest correction due to $[{\cal R},\,{\cal W}]$ 
is always two orders 
higher, while the terms from $[{\cal R},\, {\cal H}]$ 
are four orders higher, than the
similar terms in the left hand side of (\ref{sm_equation}). Their contributions
become less important \cite{lim}
because there the small statistical weight $\propto 1/\sqrt{\Omega}$ of the
collective mode is not sufficiently compensated by the inverse powers of the
low
frequency $\omega$. We can note parenthetically that this compensation can
occur only in finite systems so that the whole approach is tailored for
mesoscopic physics. Since the typical
estimates for the realistic soft modes show the dominance of the quartic
anharmonicity, we keep the main corrections to the RPA terms
\begin{equation}
\delta^{(1\,0)}={1\over 2}\left (\left [w^{(2\,0)},\,r^{(0\,1)}
\right ]_{+}-\left [w^{(0\,1)},\,r^{(2\,0)}
\right ]_{+}+\left [w^{(1\,0)},\,r^{(1\,1)}
\right ]_{+}-\left [w^{(1\,1)},\,r^{(1\,0)}
\right ]_{+} \right )\,,                                   \label{form_delta10}
\end{equation}
\begin{equation}
\delta^{(0\,1)}={1\over 2}\left (\left [w^{(1\,1)},\,r^{(0\,1)}
\right ]_{+}-\left [w^{(0\,1)},\,r^{(1\,1)}
\right ]_{+}+\left [w^{(1\,0)},\,r^{(0\,2)}
\right ]_{+}-\left [w^{(0\,2)},\,r^{(1\,0)}
\right ]_{+} \right )\,.                              \label{form_delta01}
\end{equation} 
Let us stress here that the method suggested above differs from numerous
attempts at boson expansion, see \cite{KM} and references therein. We do not
map the wave functions from the microscopic space to a bosonic one. We also do
not map directly the operators of observables. We map the {\sl equations of
motion} explicitly truncating the intermediate states. This method is regular,
does not violate general principles, and, being in fact variational, allows 
for the further improvements by including other intermediate states.

\section{Lipkin-Meshkov-Glick model} 

Using the commonly accepted procedure of testing the validity of many-body 
approximation techniques, we apply the method to 
the two-level Lipkin-Meshkov-Glick (LMG) model \cite{lipkin65}.
The space contains two single-particle levels of energies $\pm \epsilon /2$ 
with a large degeneracy $\Omega/2$ of each of them. 
We label the $\Omega$ fermionic states by quantum numbers 
$(\sigma l)\,,$ where $\sigma=\pm 1$ denotes one of the two single-particle  
levels and $l=1,2,\ldots ,\Omega/2$ distinguishes the degenerate states on each
orbital. The many-body Hamiltonian of the system is 
\begin{equation}
H={\epsilon\over 2} \sum_{\sigma,\,l} \sigma a^{\dagger}_{\sigma,\,l} \, 
a_{\sigma,\,l}\,+\, {1\over 2}V \sum_{\sigma,\,l,\, l^{\prime}} 
a^{\dagger}_{\sigma,\,l}\,a^{\dagger}_{\sigma,\,l^{\prime}}\,
a_{-\sigma,\,l^{\prime}}\,a_{-\sigma,\,l}\,.
                                                        \label{generalH}   
\end{equation}
The special feature of the problem is that the collective dynamics are
expressed in terms of the quasimomentum operators $J_{\pm}, J_{z}$,
\begin{equation}
J_{+}=J_{-}^{\dagger}=J_{x}+iJ_{y}=\sum_{l} a^{\dagger}_{+1,\,l}\, 
a_{-1,\,l}, \quad J_{z}={1\over 2} \sum_{\sigma,\,l} \sigma 
a^{\dagger}_{\sigma,\,l} \, a_{\sigma,\,l}.                  \label{e8}
\end{equation}                                     
The Hamiltonian (\ref{generalH}) can be expressed as
\begin{equation}
H=\epsilon J_{z} + {1\over 2} V (J_{+}^2+J_{-}^2)= \epsilon J_{z} + 
V (J_{x}^2-J_{y}^2)\,.                                \label{hamiltonianj}
\end{equation}

The LGM model is ideally suited to our approximate mapping procedure.
The SU(2) symmetry of the problem can be also combined with particle-hole
symmetry, which allows us to limit the consideration to the cases with the 
particle number $N\leq \Omega/2$, 
and discrete symmetries (we can take $V>0$ without loss
of generality). For the unperturbed system, $V=0$, the ground state with all
$N$ particles on the lower level belongs to the largest representation 
$J=J_{z}=N/2$ with ${\bf J}^2=[(N/2)+1](N/2)$, and then $J_{+}$ is 
an operator that creates a collective state. In this model the collective
degrees of freedom are decoupled exactly, and we need to reproduce 
the equations of motion 
\begin{equation}
[J_{x},\,H] =-i \epsilon J_{y}- i V [J_{y},\,J_{z}]_{+}, \quad
[J_{y},\,H] =i \epsilon J_{x}- i V [J_{x},\,J_{z}]_{+} \,,  \label{e9}
\end{equation}
\begin{equation}
[J_{z},\,H] = 2i V [J_{x},\,J_{y}]_{+} \,,                  \label{e10}
\end{equation}
in the mapped space of collective variables $\alpha\,,$ $\pi\,,$ with a 
collective anharmonic Hamiltonian (\ref{sm_ham1}). The kinematic constraints, 
analogous to eq. (\ref{form_algebra}, arise from the mapping of the 
quasimomentum  algebra onto a Heisenberg algebra of $\alpha$ and $\pi$.
These constraints can be accounted for by the Holstein-Primakoff transformation
\begin{equation}
J_{+}=J_{-}^{\dagger}={\cal A}^{\dagger} \sqrt{2J- {\cal A}^{\dagger} 
{\cal A}}\,=\, \sqrt{2J+1 - {\cal A}^{\dagger} {\cal A}}\;{\cal A}^{\dagger},
\quad J_{z}=-J+{\cal A}^{\dagger} {\cal A}\,, 
                                                  \label{holstein40}
\end{equation} 
where operators ${\cal A}$ and ${\cal A}^{\dagger}$ are bosonic annihilation
and creation operators, $[{\cal A},\,{\cal A}^{\dagger}]=1$.

The RPA corresponds to keeping only quadratic terms in ${\cal A}$ and
${\cal A}^{\dagger}$ which leads to the RPA Hamiltonian
\begin{equation}
H_{\rm RPA}=-\epsilon\, \left(J+{1\over2}\right)+{\epsilon \over 2} 
({\cal A}^{\dagger} {\cal A}+{\cal A}{\cal A}^{\dagger})+ 
{V\over 4}{\sqrt{16\,J^2 - 1}}  
\left (({\cal A}^{\dagger})^2 + {\cal A}^2 \right )\,.      \label{e11}
\end{equation}
The diagonalization of (\ref{e11}) results in the harmonic approximation with 
an RPA frequency \cite{meshkov65}
\begin{equation}
\omega_{\rm RPA}^2=\epsilon^2-V^2 \left (4J^2-{1\over 4} \right ). \label{e12}
\end{equation}
The instability point, $V^{2}\approx \epsilon^{2}/(4J^{2})$, exists in all
$J$-subspaces emerging first for the largest $J$ with a greater degree of
collectivity. 

The collective coordinate and momentum can be introduced with the aid of the
canonical transformation
\begin{equation}
{\cal A}={1\over \sqrt{2}} \left ( i u \alpha + v \pi \right ), \quad 
{\cal A}^{\dagger}={1\over \sqrt{2}} \left ( - i u \alpha + v \pi \right ),
\quad uv=-1.                                                     \label{uv}
\end{equation}
The LMG model has only
even order anharmonicities, and in our choice of expansion the correction 
to the $n$-th order will come from the $(n+2)$-th order in $\alpha$ and $\pi$.
An expansion up to the sixth order retaining only quadratic and quartic 
terms is necessary for determining the effective quartic Hamiltonian.
The appropriate choice of $u$ and $v$ as
\begin{equation}
u=\left(\epsilon+V {16J^2+8J-1 \over 2 \sqrt{16\,J^2 - 1}}\right)^{1/2},
\quad v=-\frac{1}{u},                                 \label{e12a}      
\end{equation} 
sets a scale of the collective Hamiltonian at $B=1$ and the parameters in 
eq. (\ref{sm_ham1}) as
\begin{equation}
\omega^2=C=\epsilon^2-V^2 { (16 J^2 + 8 J -1)^2 \over 4 (16\,J^2 - 1) }\approx
\epsilon^2-4V^2J^2\,,                                     \label{e13}
\end{equation}
\begin{equation}
\Lambda^{(4\,0)}=2V\left ({ 32 J^3 -2J -1 \over 
(16 J^2-1)^{3/2}}\right ) u^4\approx V\, u^4, \quad 
\Lambda^{(0\,4)}=-2V\left ({ 32 J^3 -2J -1 \over 
(16 J^2-1)^{3/2}}\right ) v^4\approx -V\,v^4\,.            \label{e14}
\end{equation}
At the instability point of $\omega\rightarrow 0$, assuming that $J\gg 1$,
$\epsilon \gg V$, we obtain an approximate collective Hamiltonian $(V>0)$
\begin{equation}
H={\pi^2\over 2}+ 4 \epsilon^4 \, V\, \alpha^4 \,.            \label{e15}
\end{equation}
The negative $\pi^4$ term in the collective Hamiltonian is very small in the
vicinity of the instability point in contrast to the quartic potential 
$\alpha^{4}$ which
has large matrix elements because of the large amplitude of collective
motion. This is a typical situation which emerges with a suitable choice
of collective coordinates (T-even) and collective momenta (T-odd). The next
order terms and, in general, coupling to non-collective space,
will correct the behavior of the $\pi^{4}$ term but this is 
not important for low-lying states. 

\begin{figure}
\begin{center}
\epsfxsize=14.0cm \epsfbox{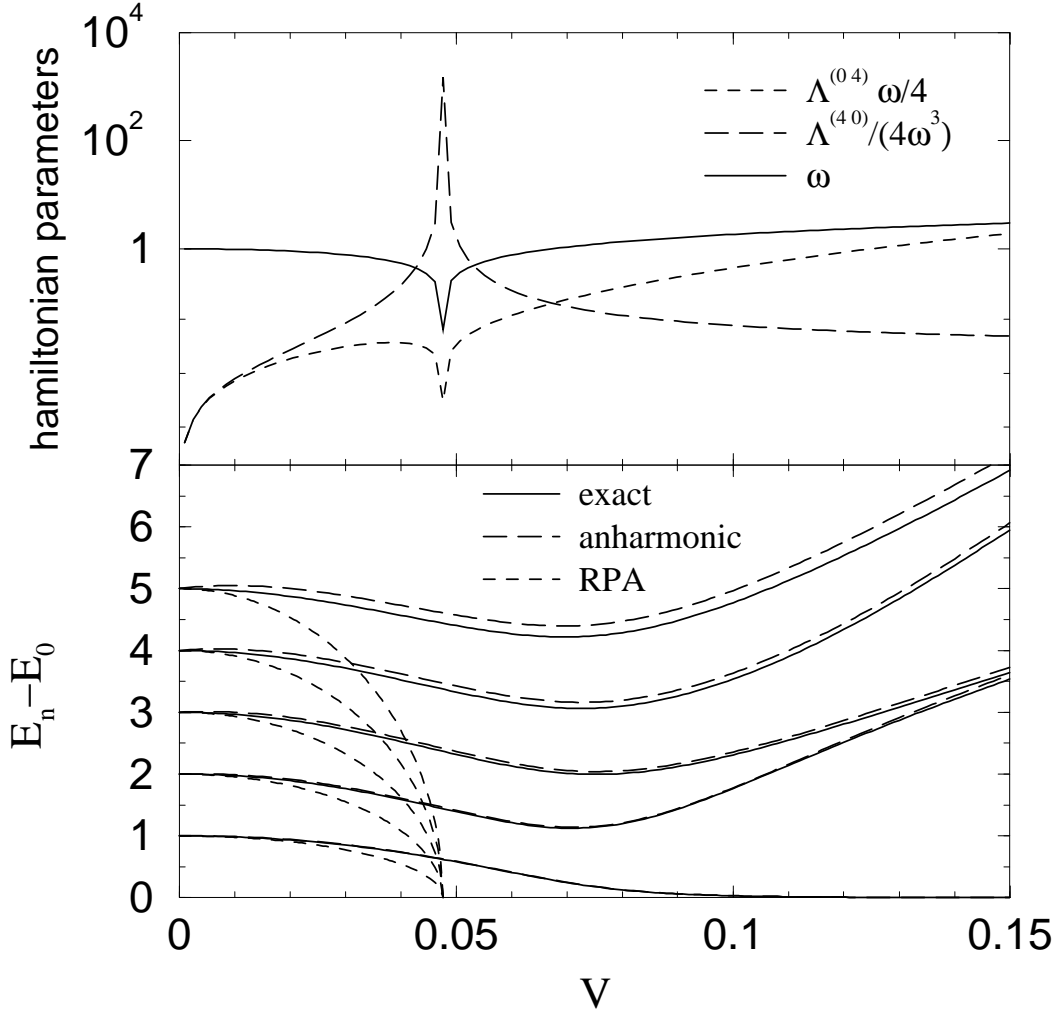}
\end{center}
\caption{
Parameters of the collective Hamiltonian (harmonic frequency and
quartic parameters) for the LGM model with 
$N=20$ particles in the largest representation $J=10$ 
as a function of the interaction strength $V$ ($\epsilon=1$), upper panel (note
the logarithmic scale with the strong growth of $\Lambda^{(40)}$ at the point of
the RPA instability); excitation energies of the first five states for the
same case of the LMG, exact solution, solid lines; RPA solution, dotted
lines; anharmonic oscillator solution with the $\pi^4$ term ignored, dashed 
lines, lower panel.   
\label{lipkinn}}
\end{figure}     

In Fig. \ref{lipkinn} we show the behavior of the harmonic term $(\omega)$
and quartic corrections $\Lambda^{(4 0)}$ and $\Lambda^{(0 4)}$ in the
dimensionless normalization, upper panel, and present a comparison of the 
exact LMG model spectrum (solid lines), RPA solution (dotted lines), and an  
improved anharmonic oscillator solution with the ignored divergent $\pi^4$
part (at large $V$ it should be included along with the high order coordinate
terms), lower panel. The anharmonic effects produce a
dramatic improvement as compared to the RPA. At the point of the RPA
instability, the contribution from the large quartic potential
restores the stability.
As the interaction strength $V$ increases, the effective potential 
\begin{equation}
U(\alpha)=\frac{1}{2}\omega^{2}\alpha^{2}+\frac{1}{4}\Lambda^{(40)}\alpha^{4}
                                                             \label{U}
\end{equation}
evolves, Fig. 2, from the harmonic oscillator to a pure quartic oscillator
at the instability point, and to the ``Mexican hat" potential with two minima.
In the last limit the lowest states of opposite parities located in the minima 
become degenerate as clearly seen in Fig. 2 (``chiral symmetry"). Contrary to
the macroscopic second order phase transition, the higher states are located
above the barrier and feel only the main quartic potential.
A similar
phenomenon should exist in soft nuclei beyond the RPA instability point when
only the lowest states are influenced by the presence of the minima in the
$\beta$ coordinate; however, there the minima are connected along the $\gamma$
coordinate which is absent in the LGM.
\begin{figure}
\begin{center}
\epsfxsize=14.0cm \epsfbox{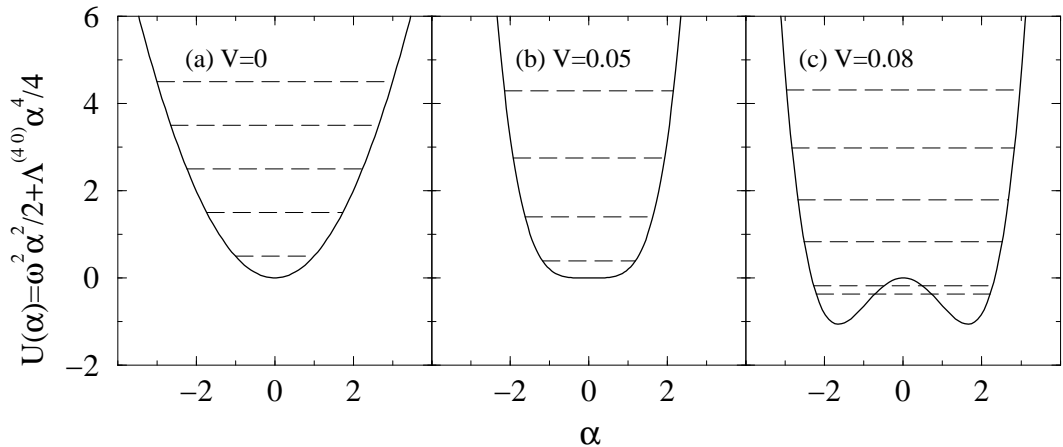}
\end{center}
\caption{The spectrum of the lowest levels in the model of Fig. 1, dashed
lines, and the shape of the effective potential $U(\alpha)$ for (a) $V=0$, the
harmonic limit; (b) $V=0.05\,,$ pure quartic potential at the RPA instability
point; (c) $V=0.08$, beyond the instability point, the splitting of the levels
of opposite parity (symmetry in $J_{z}$ in the limit of $VJ^{2}\gg \epsilon$)
below the barrier decreases as $V$ increases.  
\label{potential}}
\end{figure}

\section{Conclusion}

We have suggested an alternative approach to the construction of a collective
Hamiltonian for large amplitude collective motion in a finite
Fermi system in the presence of a soft vibrational mode. In such a
situation, the RPA is insufficient as near the RPA instability the
anharmonic effects dominate. The advantages compared to conventional 
techniques, such as the generator coordinate method, are related to a fully
quantum consideration which does not require the derivation of an approximate
classical Hamiltonian with the subsequent ill-defined procedure of
requantization. It differs from the approaches utilizing various versions
of the boson expansion in the variational character of the formalism. We can
vary the collective space assuming the saturation of the exact operator
equations of motion within a part of total Hilbert space. With our techniques
we also avoid the slow convergence problem of the naive boson expansion.

Of course, the illustrative example of the Lipkin model is perfectly suited to
our goal since the collective modes of this model are fully decoupled. 
However it emphasizes the predominance of the quartic anharmonicity near the RPA
instability. Because of the convenient operator distinction between the
coordinate and momentum parts, we concentrate the most important anharmonic
effects in the quartic potential which has large matrix elements in the
dangerous region of interest. 

In realistic cases, the
collective space is not decoupled completely. The effects
of coupling to noncollective states lead to the spreading of the collective
strength and the chaotization of motion in the region of high level density.
To treat this situation as well, we can include the matrix elements of the GDM
connecting the collective band with incoherent states. One promising approach
would be to consider these states on average, making the random phase assumption
on a new level of treatment. This would introduce an effective background for
the collective mode to describe its spreading and damping.\\
\\
{\small The authors are thankful to D. Mulhall for assistance and criticism;
they acknowledge support from the USA National Science 
Foundation.}


\begin{references}
\bibitem{bel} S.T. Belyaev, Mat. Fys. Medd. Dan. Vid. Selsk. {\bf 31}, No. 11
(1959).
\bibitem{BE} S.T. Belyaev and V.G. Zelevinsky, Nucl. Phys. {\bf 39} (1962) 582.
\bibitem{ring} P. Ring and P. Schuck, {\sl The Nuclear Many-Body Problem}
(Springer, Berlin - New York, 2000).
\bibitem{solov} V.G. Soloviev, {\sl Theory of Atomic Nuclei; Quasiparticles and
Phonons} (Institute of Physics, Bristol - Philadelphia, 1992).
\bibitem{KM} A. Klein and E.R. Marshalek, Rev. Mod. Phys. {\bf 63} (1991) 376.
\bibitem{lim} V.G. Zelevinsky, Int. J. Mod. Phys. {\bf E2} (1993) 273.
\bibitem{gamm} M. Jean and L. Wilets, Phys. Rev. {\bf 102} (1956) 788.
\bibitem{IBM} F. Iachello and A. Arima, {\sl The Interacting Boson Model}
(University Press, Cambridge, 1987).
\bibitem{quar} O.K. Vorov and V.G. Zelevinsky, Sov. J. Nucl. Phys. {\bf 37}
(1983) 830.
\bibitem{VZ} O.K. Vorov and V.G. Zelevinsky, Nucl. Phys. {\bf A439} (1985) 207.
\bibitem{arm} J. Armstrong and V. Zelevinsky, Preprint MSUCL-1098, April 1998;
BAPS {\bf 44}, Part I, p. 397.
\bibitem{iach} F. Iachello, Phys. Rev. Lett. {\bf 85} (2000) 3580.
\bibitem{sol} V.G. Zelevinsky, in {\sl Frontiers in Nuclear Physics} (Dubna,
1995) p. 243.
\bibitem{denmat} S.T. Belyaev and V.G. Zelevinsky, Sov. J. Nucl. Phys. {\bf 16}
(1973) 657.
\bibitem{nucold} V.G. Zelevinsky, Nucl. Phys. {\bf A337} (1980) 40.
\bibitem{supp} V.G. Zelevinsky, Prog. Theor. Phys. Suppl. {\bf 74-75} (1983)
251.
\bibitem{MF}. V.G. Zelevinsky, Nucl. Phys. {\bf A555} (1993) 109.
\bibitem{jolos} R.V. Jolos, P. von Brentano, N. Pietralla, and I. Schneiden,
Nucl. Phys. {\bf A618} (1997) 126.
\bibitem{lipkin65} H. Lipkin, N. Meshkov, and A. Glick, Nucl. Phys.
{\bf 62}, (1965) 188.                      
\bibitem{meshkov65} N. Meshkov, A. Glick, and H. Lipkin, Nucl. Phys. {\bf 62}
(1965) 199.
\end{references}
\end{document}